\documentstyle[12pt,aps,preprint,psfig]{revtex}
\draft
\tightenlines
\begin{document}
\newcommand{\Trace}{\mbox{Trace}}
\newcommand{\rr}{{\bf r}}
\newcommand{\p}{{\bf p}}
\newcommand{\k}{{\bf k}}
\newcommand{\R}{{\bf R}}
\newcommand{\PP}{{\bf P}}
\newcommand{\x}{{\bf x}}
\newcommand{\be}{\begin{equation}}
\newcommand{\ee}{\end{equation}}
\newcommand{\bea}{\begin{eqnarray}}
\newcommand{\eea}{\end{eqnarray}}
\newcommand{\f}{\frac}  
\newcommand{\pa}{\partial}
\newcommand{\la}{\lambda}
\newcommand{\ve}{\varepsilon}
\newcommand{\ep}{\epsilon}
\newcommand{\D}{{\bf D}}
\newcommand{\da}{\downarrow}
\newcommand{\V}{{\cal V}}

\title{Saturation of the width of the strength function
\footnote{Supported in part by FAPESP.}}

\author{A.J. Sargeant, M.S. Hussein, M.P. Pato and M. Ueda}
\address{Nuclear Theory and Elementary Particle Phenomenology Group,
Instituto de F\'{i}sica, Universidade de S\~{a}o Paulo,
C.P. 66318, 05315-970 S\~{a}o Paulo, SP, Brazil}
\date{\today}
\maketitle
\begin{abstract}
The strength function of a single state $|d\rangle$ is studied using the deformed
Gaussian orthogonal ensemble. In particular we study the dependence of the spreading width of 
$|d\rangle$ on the degree of mixing.
\end{abstract}

\vspace{.5in}
The mixing of a single state with a background of complicated states is important 
in the description of a variety of phenomena in nuclear physics such as isobaric analogue
resonances, giant dipole resonances and the decay out superdeformed rotational bands.
Such can mixing be conveniently described by the strength function \cite{Bo 69} and it is 
interesting to
study the generic features of this object. Previous studies in this vein have investigated a single
state coupled to a background generated by a two-dimensional anharmonic oscillator 
\cite{Hi 97} and the spreading of a shell model basis state over the shell model eigenstates
due to the the residual interaction \cite{Fr 96}. Here we use random matrix theory \cite{Gu 98}.

We write the Hamiltonian, H, as the sum of two terms,
\be
H=H_0+\V.
\label{Ham}
\ee
The eigenstates, $|n\rangle$, and eigenvalues, $E_n$, of $H$ satisfy
\be
H|n\rangle=E_n|n\rangle,\hspace{0.2cm} n=1,...N+1,
\label{eigH}
\ee
whilst for $H_0$ we have
\bea
\nonumber
H_0|k\rangle&=&E_k|k\rangle,\hspace{0.2cm} k=1,...N,
\\
H_0|d\rangle&=&E_d|d\rangle,\hspace{0.2cm} d=N+1.
\label{eigH0}
\eea
The strength function is then defined as
\be
F_d(E)=\sum_{n=1}^{N+1}|\langle d|n\rangle|^2\delta(E-E_n),
\label{Fd}
\ee
and describes how the state $|d\rangle$ is distributed over 
the $N+1$ eigenstates of $H$. We shall choose the energy of $|d\rangle$, $E_d$, 
such that it lies in the middle of the the spectrum of $H$.

Some insight into the behaviour of the strength function can be obtained
by performing a two step diagonalisation of $H$ \cite {Bo 69}.
Let us represent the Hamiltonian in the basis $\left\{|k\rangle,|d\rangle\right\}$.
Diagonalising $H$ in the $N$--dimensional subspace defined by
excluding $|d\rangle$ we obtain the set of eigenvectors 
$|q\rangle=\sum_k\langle k|q\rangle|k\rangle$ with eigenvalues $E_q$,
$q=1,...,N$. The Hamiltonian in the basis $\left\{|q\rangle,|d\rangle\right\}$
has diagonal matrix elements $E_q$, $q=1,...,N$ and $E_d$, $d=N+1$
and non-zero off-diagonal elements
$
\V_{dq}=\V_{qd}=\sum_k\V_{dk}\langle k|q\rangle
$
(the Hamiltonian is assumed to be real symmetric).
The diagonalisation of the intermediate matrix may be carried out analytically \cite{Bo 69}
so that using a Lorentzian of width $I$ to represent the $\delta$--function in Eq. (\ref{Fd}), one
obtains \cite{Bo 69}
\be
F_d(E)=\f{1}{2\pi}\f{\Gamma^{\da}_{d}+I}{(E-E_d-\Delta^{\da}_{d})^2+(\f{\Gamma^{\da}_{d}+I}{2})^2},
\label{Fd2}
\ee
where
\bea
\Gamma^{\da}_{d}(E)=I\sum_q\f{|\V_{dq}|^2}{(E-E_q)^2+(\f{I}{2})^2}
\label{gamdda}
\eea
and
\bea
\Delta^{\da}_{d}(E)=\sum_q\f{|\V_{dq}|^2(E-E_q)}{(E-E_q)^2+(\f{I}{2})^2}.
\label{deldda}
\eea
By making the further assumptions that the eigenvalues $E_q$
are equi-distant with mean spacing $D$, that the squared matrix elements $|\V_{qd}|^2$ have approximately
the same order of magnitude, $\langle\V^2\rangle$, for all $q$ and that the magnitude of
$\sqrt{\langle\V^2\rangle}$ is smaller than the energy range in which it may be considered constant 
(whilst being larger than $D$ in order for the strength function to have meaning) the strength function
may be approximated by a Lorentzian \cite{Bo 69,Fr 96}
\be
F_d(E)\approx\f{1}{2\pi}\f{\Gamma^{\da}_{d}}{(E-E_d)^2+(\f{\Gamma^{\da}_{d}}{2})^2},
\label{Fdw}
\ee
where the spreading width is given by the ``golden rule''
\be
\Gamma^{\da}_{d}\approx 2\pi\f{\langle\V^2\rangle}{D}.
\label{Gdd}
\ee

We wish to study how the distribution, $F_d(E)$, depends
on the degree of mixing by which we mean the strength of $\V$.
To that end we employ deformed Gaussian orthogonal
ensemble (DGOE) \cite{Hu 93}. 
In this model $H$ is real symmetric and it's matrix elements
are taken to be independent Gaussian distributed random numbers
with zero mean and variances $\langle (H_0)^2_{k,k}\rangle=\f{a^2}{2N}$ for the diagonal matrix 
elements and $\langle (\V)^2_{k,k'}\rangle=\f{\la^2 a^2}{4N}$ for the off-diagonal matrix elements. 
We take $\langle (\V)^2_{k,d}\rangle=\langle (\V)^2_{d,k}\rangle=\f{\la^2 a^2}{4N}$ as well
although choosing a different variance for this matrix element may be appropriate in some 
applications.

The parameter $\la$ may be varied between $0$ and $1$ and determines the degree
of mixing.
The parameter $a$ determines the energy interval over which the
eigenvalues are distributed. In the limiting case $\la=0$ the eigenvalues $E_n$ are 
Gaussian distributed;
\be
\rho_0(E)=\f{N^{\f{3}{2}}}{a\sqrt{\pi}}\mbox{exp}(\f{-NE^2}{a^2}),
\label{denpoi}
\ee
whilst in the limit $\la=1$ they are distributed according to the Wigner semicircular law;
\be
\rho_1(E)=N\f{2}{\pi a^2}\sqrt{a^2-E^2}.
\label{dengoe}
\ee

We perform an unfolding of the spectrum of the spectrum of $H$ defined by
\be
x_n=\int_{-\infty}^{E_n}dE\rho(E),
\label{x}
\ee
where $\rho(E)$ is the smoothly varying part of the level density.
Thus the unfolded spectrum has mean level density equal to 1 and 
is dimensionless. The smooth variation the level density was
obtained in practice by fitting the cumulative level density (staircase function)
\be
x=\sum_{n=1}^{N+1}\theta(E-E_n)
\label{cum}
\ee
(the letter $\theta$ denotes the unit step function) to a polynomial using 
the method of linear least squares \cite{Pr 94}. For the case $N=50$ 
(used in the calculations below) we found
the function which gave the best visual fit was a polynomial of degree 5.

In order to compare our calculation of the strength 
function on the unfolded energy scale with the Lorentzian approximation, Eqs. (\ref{Fdw})
and (\ref{Gdd}), we apply a simplified unfolding to these approximate formulae, defined by 
\be
x=\f{E}{D}
\label{xs}.
\ee
Thus the unfolded version of the golden rule is
\be
\f{\Gamma^{\da}_{d}}{D}\approx 2\pi\f{\langle\V^2\rangle}{D^2}.
\label{Gddsx}
\ee

Persson and  \.{A}berg \cite{Pe 95} proposed the following formula for the mean
(meaning averaged over the energy range where the eigenvalues are distributed) 
density of states of the DGOE which interpolates between $\la=0$ and $\la=1$, valid when $a=2$:
\be
\bar{\rho}_{\la}=\f{N^{\f{3}{2}}}{4\la N^{\f{1}{2}}+7N^{-1.5\la}}.
\label{abden}
\ee

\begin{figure}[ht]
\centerline{\psfig{figure=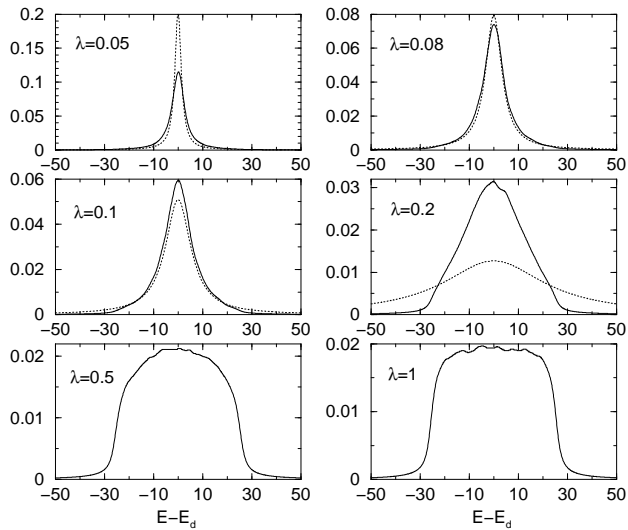,width=0.5\textwidth}}
\caption{Strength function, $F_d(E)$, for various values of the mixing parameter, $\la$,
calculated using  DGOE (solid lines) and calculated 
using the Lorentzian approximation (dotted lines). See text for discussion.}
\label{fig1v2}
\end{figure}

Figure \ref{fig1v2} displays our calcuations of the strength function (solid lines) 
for various $\la$ using Eqs. (\ref{Fd2}--\ref{deldda}) where the eigenvalues 
(unfolded using the procedure defined by Eq. (\ref{x})) and eigenvectors are generated by the DGOE. 
We performed our calculations using $N=50$ whilst the width of the the Lorentzian used 
to represent the $\delta$--function (Eq. (\ref{Fd})) is $I=3$. We set $a=2$.

The dotted lines in fig. \ref{fig1v2} are the strength function calculated using the
Lorentzian approximation, Eqs. (\ref{Fdw}) and (\ref{Gdd}). The mean square value of the
coupling is taken to be $\langle\V^2\rangle\approx\f{\la^2 a^2}{4N}$. 
The mean level spacing $D$ is taken to be
$\f{1}{\rho_0(E=0)}=\f{a\sqrt\pi}{N^{\f{3}{2}}}$; the
lower limit for the level spacing of the DGOE spectra.

Using the estimates of the previous paragraph,
the condition $\f{\langle\V^2\rangle}{D^2}>1$ (for the single state $|d\rangle$ to be significantly
mixed with more than a single one of the $|q\rangle$) implies $\la>\f{2\sqrt\pi}{N}$. 
Thus for  $N=50$, although we can calculate the strength function for arbitrary small $\la$
it is only meaningful for $\la>0.07$. The value of $I$ should be
chosen so that it is a negligable fraction of the combined width $\Gamma_d^{\da}+I$
whilst being greater than the level spacing (unity after unfolding).  

From fig. \ref{fig1v2} we can see that for weak $\la$ the strength function has an 
approximately Lorentzian shape. As we decrease $\la$ below 0.07 the strength function
approaches a Lorentzian whose width is dominated by $I$. For $\la>0.07$ the 
strength function starts to deviate significantly from the Lorentzian shape already at 
$\la=0.1$ broadening towards a semicircle before $\la=1$.  An ensemble average 
was performed over 100 realizations for the cases $\la=0.05$, 0.08 and 0.1 whilst to
obtain a relatively smooth strength function for the $\la=0.2$, 0.5 and  it was
necessary average over 1000 realizations.

\begin{figure}[ht]
\centerline{\psfig{figure=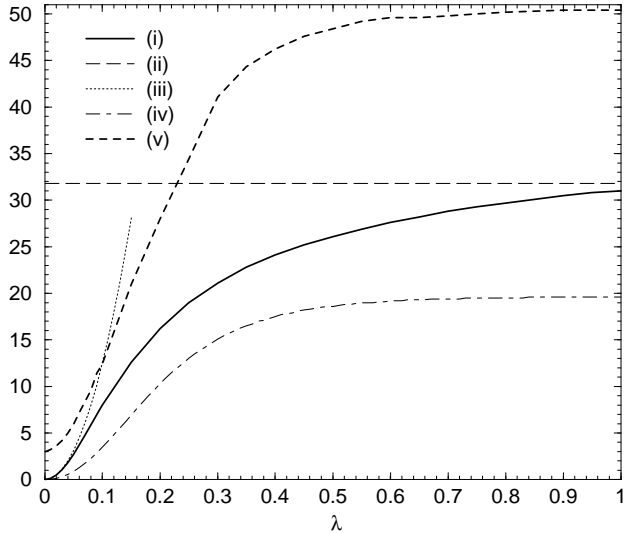,width=0.5\textwidth}}
\caption{Spreading width of $|d\rangle$ as a function of the mixing parameter $\la$. See text for
meaning of (i)--(iv).}
\label{fig2v3}
\end{figure}

In fig. \ref{fig2v3} we plot the following calculations for the spreading width of
$|d\rangle$ as a function of $\la$:
\\(i)  Eq. (\ref{gamdda}), at the peak value $E=E_d$
where the eigenvalues (unfolded using the procedure defined
by Eq. (\ref{x})) and eigenvectors are generated by the DGOE. An ensemble 
average was performed over 100 realisations for all $\la$; otherwise the same
parameters are used as were used in fig. \ref{fig1v2}.
\\(ii) The limiting value ($\la=1$) for the width calculated using the golden rule
(Eq. \ref{Gddsx}) and density Eq. (\ref{dengoe}) at $E=0$: 
$2\pi\left(\f{a^2}{4N}\right)\left(\f{2N}{a\pi}\right)^2=\f{2N}{\pi} $. 
\\(iii) The golden rule expression for the width using the 
Poisson density, Eq. (\ref{denpoi}), at $E=0$: 
$2\pi\left(\f{\la^2a^2}{4N}\right)\left(\f{N^{\f{3}{2}}}{a\sqrt{\pi}}\right)^2=\f{\la^2N^2}{2}$.
\\(iv) The golden rule with the density Eq. (\ref{abden}):  
$2\pi\left(\f{\la^2a^2}{4N}\right)\bar{\rho}_{\la}^2$.
\\(v) The full width at half maximum FWHM of the strength function $F_d(E)$, 
Eqs. (\ref{Fd2}--\ref{deldda}), calculated using the DGOE.

Three regions may be identified in the DGOE calculations (lines (i) and (v))
in fig. \ref{fig2v3}). We see that $\Gamma_d^{\da}(E_d)$ for very weak coupling 
($\la<0.04$) has a quadratic dependence which accurately follows the
golden rule calculation (line (ii)).  Between $\la=0.05$ and $\la=0.15$ the 
this dependence is linear becoming very weak as $\la$ approaches unity.  
The behaviour of the DGOE is simulated by Eq. (\ref{abden}) 
of Persson and \.{A}berg for the average level density (line (iv)). 
At $\la=1$ our DGOE calculation of $\Gamma_d^{\da}(E_d)$ is approximately a factor 
$\left(\f{\rho_1(E=0)}{\bar\rho_1}\right)^2=\left(\f{4}{\pi}\right)^2=1.6$ greater than 
the calculation which employs Eq. (\ref{abden}) .
The FWHM (line (v)) also has a quadratic 
dependence on $\la$ for weak $\la$. A linear dependence is maintained
up to $\la=0.3$ after which the $\la$--dependence becomes very weak as 
$\la$ approaches unity. When $\la=0$ the FWHM is just equal to the 
averaging interval $I(=3)$. In the opposite limit of $\la=1$ the FWHM is
essentially equal to $N(=50)$; ie. the state $|d\rangle$ is spread over all the 
eigenstates of $H$.
Thus in the limit $\la=1$, the golden rule (line (v)) is 
a factor $\f{2}{\pi}$ smaller than the FWHM.

Ref. \cite{Fr 96} investigated the spreading width of a basis state of a shell model Hamiltonian
as a function of the strength of the residual interaction (corresponding to $\la a$ above). 
They identified regions where the spreading width has 
a quadratic dependence on the $\la$ (weak mixing) becoming linear for larger $\la$.
They also found that the golden rule cannot be used to estimate the FWHM of the strength
function for strong mixing.

In conclusion we have studied how the shape and width of the the strength function
of a single state depends on the degree of mixing,
using random matrix theory.  An application of these results to
the decay out of a superdeformed rotational band is in progress.

A.J.S. acknowledges helpful comments on the manuscript by Professor R.C. Johnson.

\end{document}